August 31, 2020

# Stability analysis for cylindrical Couette flow of compressible fluids

Christian Fronsdal

*Dept. of Physics and Astronomy, University of California Los Angeles, USA*

*ABSTRACT* A new analysis of basic Couette flow, is based on an Action Principle for compressible fluids, with a Hamiltonian as well as a kinetic potential. An effective criterion for stability recognizes the tensile strength of water. This interpretation relates the problem to capillary action and to metastable configurations (Berthelot's negative pressure experiment of 1850). We calculate the pressure and density profiles and find that the first instability of basic Couette flow is localized near the bubble point. This theoretical prediction has been confirmed by recent experiments.

The theory is the result of merging the two versions of classical hydrodynamics, as advocated by Landau for superfluid Helium II. In an inspired paper, Landau (1941) introduced the idea of two independent flows, 'phonons' and 'rotons", with strong emphasis on the idea that <u>there is only one kind</u> <u>of fluid.</u> The dynamical variables include two flows but only one density variable.

In this paper two-flow dynamics is created by merging two actions, not by choosing between them, nor by combining the two vector fields as in the Navier-Stokes equation. At rest, as contributions to the mass flow they cancel, but a non-zero kinetic energy and kinetic potential as well as non-zero angular momentum remain, manifest as liquid tension, as is well known to exist by observation of the meniscus and configurations with negative pressure. (Fronsdal 2020b in preparation).

The immediate effect of merging the two versions of classical hydrodynamics, in a unique theory based on an action principle, is to provide a Hamiltonian and a kinetic potential for compressible fluids with rotational flow. This theory gives a very satisfactory characterization of the limit of stability of the most basic Couette flow. The inclusion of a vector field that is not a gradient has the additional affect of introducing spin, which explains a most perplexing experimental discovery: the ability of frozen Helium to remember its angular momentum (spin).

## I. Introduction. Mixed flows in hydrodynamics

An action principle due to Lagrange (1760) defines the 'Eulerian version' of hydrodynamics. The Galilei invariant action is

$$A_1[\rho,\Phi] = \int dtd^3x \,\Big(\rho(\dot\Phi - \vec\bigtriangledown\Phi^2/2 - \varphi) - W[\rho]\Big)\,. \tag{1.1}$$

Gravitation is represented by the static field $\varphi$. The velocity is irrotational,

$$\vec u = -\vec\bigtriangledown\Phi, \tag{1.2}$$

which severely limits the fields of applications. Efficiency and elegance has made Lagrange's theory (Lagrange 1781) extremely popular but the exclusion of vorticity has prevented successful applications to important problems such as lift and drag (Prandtl 1914, Eckert 2006), to ocean waves (Lamb 1932, Munk and Sverdrup 1943). Among attempts to lift the theory to a more general context, some have come close to formulating a more general action principle. This paper follows a direction suggested in the context of superfluid Helium by Landau (1941). But to get to the heart of the matter we should go further back.

The 19'th century saw an intense search for action principles for thermodynamics, by Helmholtz, Gibbs, Maxwell, Poincaré and others. Einstein's great achievement of 1915 was the discovery of the action for a theory of gravity. The spectacular advance of particle theory during the 20'th century was based on Lagrangian variational principles. In a memorandum to O. Veblen, dated March 26, 1945, John von Neuman (1945) laments the fact that "hydrodynamical problems, which ought to be considered relatively simple, offer altogether disproportionate difficulties"; he says that "the true technical reason appears to be that variational methods have ... hardly been introduced in hydrodynamics." The reason that they had not been used is to some extent explained by the internal inconsistencies related to the use (or misuse) of the energy concept. Actually, the attraction of canonical methods had remained very strong, as is evidenced by the fact that the shortcomings of the Eulerian theory were viewed as 'paradoxical' (Birkhoff 1950).

This paper adopts as a premise that the search for an action principle for hydrodynamics (and thermodynamics) was inspired and that it is ultimately headed for success. That difficulties were encountered only reinforces the expectation that the rewards will be a unique theory with very strong powers of prediction, the ultimate goal of any physical theory. Let us take a moment to consider this issue.

Traditional hydrodynamics has 4 independent field variables: the density and the 3 components of the velocity. The action (1.1) has only 2 independent variables, the density and the velocity potential; confirming that more are needed. But we shall show that it will not do to relax the formula (1.2).

The restriction (1.2) to vortex-less flows is certainly too strong, but the scalar velocity potential is nevertheless essential. The Euler-Lagrange equation

$$\delta A_1 = \int dtd^3x \; \delta\Phi \left(\dot\rho + \vec\bigtriangledown \cdot (\rho\vec v)\right) = 0 \tag{1.3}$$

is the equation of continuity, the very essence of hydrodynamics. Additional considerations (Appendix) will show that we must add more variables, but retain the scalar velocity potential and the equation of continuity intact.

## Two types of flow

Irrotational flows are familiar, but other types of flow occur as well, among them are rotating fluids that maintain a fixed profile, like a solid body:

$$\dot{\vec X} = \omega(-y, x, 0), \quad \omega = \text{constant}.$$

These too can be described by an action principle, with the action

$$A_2[\rho, X] = \int dtd^3x \; \rho\Big(\dot{\vec X}^2/2 - W[\rho]\Big).$$

This type of vector field was used to describe 'rotons' in Landau's theory of superfluids (Landau 1941). In that context it was noticed (Lund and Regge 1964) that the theory, to be unitary, has to be interpreted as a gauge theory in a special gauge. The full gauge invariant theory is that of a relativistic 2-form, related to $\vec X$,

$$Y_{jk} = X^i\epsilon_{ijk},$$

while $Y_{0i} = 0$ in the chosen gauge.

Long before this, Cauchy developed the concept of a stress tensor that summarizes the conservation laws of a theory invariant under coordinate transformations. From Cauchy's work came the Navier-Stokes equation. When this equation was applied to basic Couette flows it was noticed that the equation is solved by linear combinations

$$\vec{u} = \frac{a}{r^2}(-y, x, 0) + b(-y, x, 0). \tag{1.4}$$

But no action principle was used and it was not pointed out that the two terms are solutions of two different theories, with different transformation properties. In the non relativistic context; the 'Lagrangian' velocity field $\vec{X}$ is inert (up to a gauge transformation) under Galilei transformations. It implies that the two types of velocity must appear independently and this too rules out the possibility of replacing both by their sum. Instead we shall merge the two theories and find an interpretation of two velocities in hydrodynamics, as foreseen by Landau.

The connection of irrotational flow to an action principle results from the fact that the velocity is a gradient of a more basic field, the scalar potential. The velocity field of the alternative, 'Lagrangian' formulation of hydrodynamics, here denoted $\dot{\vec{X}}$, is the time derivative of a vector potential, and this vector potential is another natural candidate for a canonical velocity variable. Since the equations of motion are second order in the time derivative, this field appears to have 6 independent degrees of freedom. We already have 2 degrees of freedom with $\rho$ and $\Phi$, while conventional hydrodynamics has 4. It is possible, however, and necessary, to impose constraints that limit the number of degrees of freedom of the new field to just 2. [1] The constraints lead to the non-relativistic limit of the relativistic notoph gauge theory (Ogievetskij and Polubarinov 1964), with its single propagating mode.

The need to include two widely different types of vector fields in a more general type of hydrodynamics resonates strongly with Landau' two-flow hypothesis. Landau's advice (2 flows but only one density) has not always been followed in the original context of superfluids, but it has led to decisive progress in our understanding of metastable liquids, capillary action, superfluidity and General Relativity.

The complete, non-relativistic Lagrangian density of the proposed Conser-

[1] In fact, it is necessary, for unitarity.

vative Hydrodynamics is, in a fixed gauge in which $\vec\bigtriangledown\wedge\dot{\vec X}=0$, [2]

$$\mathcal{L}[\rho,\Phi,\vec X]=\rho(\dot\Phi-K-\varphi)-W[\rho], \tag{1.5}$$

$$K[\rho,\Phi,\vec X]:=-\dot{\vec X}^2/2-\kappa\rho\dot{\vec X}\cdot\vec\bigtriangledown\Phi+(\vec\bigtriangledown\Phi)^2/2. \tag{1.6}$$

The first term in $K$ accounts for the tension that is seen in the experiments with negative pressures (Berthelot experiment). The $\kappa$ term is a spin-orbit coupling that accounts for one of the most remarkable discoveries about superfluid Helium (Kim and Chan 1950). In the present context it brings in vorticity, as shown below. It is the only free parameter in a theory that is otherwise unique except that the boundary conditions that would complete it are only beginning to become known. (Fronsdal 2020b).

Variation of the action with respect to the variable $\rho$ gives the Bernoulli equation in integrated form, to be examined in Section III:

$$\dot\Phi-K-\varphi=\mu+ST=\partial W[\rho]/\partial\rho, \tag{1.7}$$

where $\mu$ is the chemical potential and $S$ is the specific entropy density. The last equality comes from thermodynamics after eliminating the temperature with the help of the adiabatic condition. The kinetic potential $K$ was defined in Eq.(1.6).

Variation of (1.5) with respect to the velocity potential leads to the continuity equation for $\rho$ and the flow velocity $\vec v$:

$$\dot\rho+\frac{d}{dt}(\rho\vec v)=0,\quad \vec v:=\kappa\dot{\vec X}-\vec\bigtriangledown\Phi. \tag{1.8}$$

The vorticity field is

$$\vec\bigtriangledown\wedge\vec v=\kappa\vec\bigtriangledown\wedge\vec w=\kappa\vec\bigtriangledown\frac{1}{\rho}\wedge\vec m, \tag{1.9}$$

where

$$\vec w=\dot{\vec X}+\kappa\vec\bigtriangledown\Phi,\quad \vec m=\rho\vec w=-\vec\bigtriangledown\tau; \tag{1.9bis}$$

it vanishes in the case that the parameter $\kappa=0$. It is orthogonal to the density gradient:

$$\vec\bigtriangledown\rho\cdot(\vec\bigtriangledown\wedge\vec v)=0, \tag{1.10}$$

[2]The complete Lagrangian includes a boundary term.

An incompressible fluid can be understood only as the limit of a theory of compressible fluids. Supporting this statement is the fact that there can be no relativistic theory of incompressible fluids.

The proposed Lagrangian, Eq.(1.5), has much in common with modern attempts to deal with rotational flow; they all involve a potental flow and an additional, constrained velocity field (Hall and Vinen 1957, Fetter 2009).

Classic stability analysis of cylindrical Couette flow, using an *ad hoc* kinetic energy and an *ad hoc* kinetic potential (Rayleigh 1916), is in contradiction with observation. The approach is repeated in many textbooks and reviews up to 2017, and its failure to agree with observation was frequently noted. More successful approaches have been found, but the reason why Rayleigh's unsuccessful stability criterion is of little significance has only recently been pointed out. A discussion should therefore prove useful.

A new analysis of Couette flow, using a fully dynamical theory of hydrodynamics with equations of motion derived from a unique action principle, presented in Section III-V and conclusions are offered in Section VI.

## II. Cylindrical Couette flow

Couette flow has been been studied for 140 years, with focus on the instabilities of the various observed flows. A homogeneous fluid is contained in the space between two concentric cylinders, as in Fig.s 1 and 2, both cylinders turning independently.

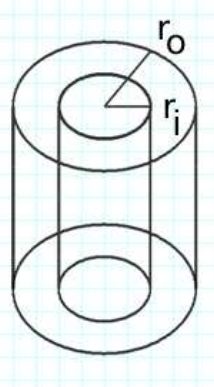

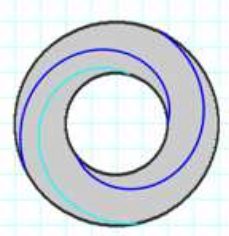

Fig.1. Cylindrical Couette flow, both cylinders turning independently.

Fig.2. Illustration of Cylindrical Couette flow. The lines are successive snap shots of a set of particles originally in a radial configuration.

A pioneering series of experiments, by Couette (1887-1890) and Mallock (1888) was followed by a work by G.I. Taylor (1923). The latter reported new experiments as well as a penetrating analysis, the main impact of which was the vindication of the 'no slip' boundary condition. [3]

Experiments start with both cylinders at rest, then they are put into independent rotational motion about the common axis. The angular speeds $\omega_i$ and $\omega_o$ are increased slowly, until instability of the flow is observed.

The most interesting presentations of experimental results show a partition of the plane of angular speeds into a stable and an unstable region (Fig.4). A critical line, approximately hyperbolic, marks the onset of instability, the stable region lies below. A completely satisfactory theoretical treatment has yet to be found (Lin 1955, Betchov and Criminale 1967, Joseph 1976, Drazin and Ried 1981, Schmid and Henningson 2000).

**Solid body flow**

The simplest type of flow velocity is stationary, horizontal and circular: in Cartesian coordinates,

$$\dot{\vec{X}} = \omega(r)(-y, x, 0), \qquad r := \sqrt{x^2 + y^2}.$$

The boundary conditions are non-slip. To simplify the analysis one thinks of the cylinders as being long, and ignore end effects. According to Andereck *et al* (1986) the aspect ratio (length to radius) does not have an important effect.

The particular case

$$\dot{\vec{X}} = b(-y, x, 0), \qquad b \text{ constant}, \tag{2.1}$$

imitates the flow of a solid body. [4] It gives rise to a kinetic potential

$$K = -\dot{\vec{X}}^2/2 \propto -r^2. \tag{2.2}$$

All proportionalities imply (by definition) positive coefficients, the sign in this equation is significant. A centrifugal force $-\vec{\bigtriangledown} K \propto \vec{r}$ is balanced by a pressure ($\vec{\bigtriangledown} p \propto \vec{\bigtriangledown} r$) that presents a force - $\vec{\bigtriangledown} p \propto -\vec{r}$. It follows, for ordinary fluids with a positive adiabatic derivative $dp/d\rho$, if there are no external forces acting, that $\rho\vec{\bigtriangledown} K + \vec{\bigtriangledown} p = 0$ and that the density must increase outwards. See Fig 3.

Irrotational flow in the same geometry is the velocity field

$$-\vec{\bigtriangledown}\Phi = \frac{a}{r^2}(-y, x, 0), \qquad r > 0.$$

[3]It is the statement that the limit of the material flow velocity, at the walls of the containing vessel, is the same as the local velocity of the container at the same point.

[4]This velocity field is not a gradient. To give substance to the claim that this formula defines a kinetic potential one must specify the action principle used and it can only be the Lagrangian version, where in fact (2.2) holds. Both types of vector fields are present!

The energy density $a^2/2r^2$ is positive and so is the the kinetic potential.

**The kinetic potential is $+\vec{v}^2/2$ for irrotational flow and $-\vec{v}^2/2$ for solid-body flow. The two kinds of flow are two different kinds of vector variables.**

The difference in sign comes from the Legendre transformation; the irrotational velocity is a spatial derivative; the solid flow velocity is a time derivative. The two flows are solutions of two different theories.

Key question: what is the expression for the 'kinetic potential' when the velocity is a sum of two parts, one irrotational and the other solid-body type?

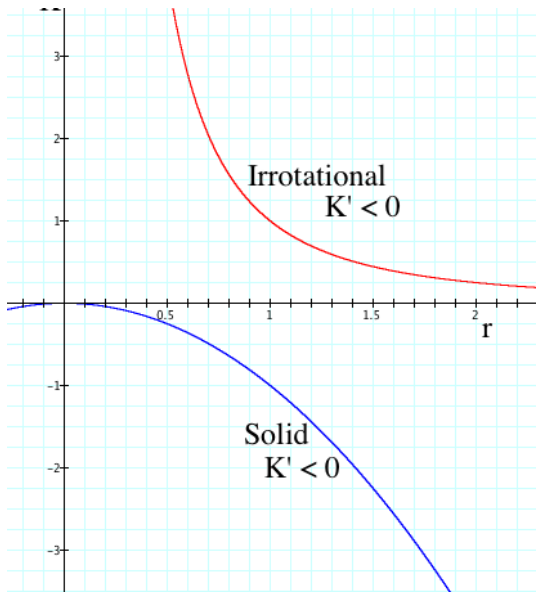

Fig. 3. The kinetic potentials, $K(r) = +\vec{v}^2/2$ for irrotational flow (see Eq.(2.6)) and $K(r) = -\vec{v}^2/2$ (for solid-body flow Eq(2.2)), both decreasing outwards as observed.

### The Navier-Stokes equation

has served as the basis for every traditional analysis of Couette flow:

$$\frac{D}{Dt}\vec{u} + \vec{\bigtriangledown}\varphi = -\frac{1}{\rho}\vec{\bigtriangledown}p + \bar{\mu}\Delta\vec{u}. \tag{2.3}$$

Here $\bar{\mu}$ is the kinematical viscosity of the fluid, in the simplest case a constant parameter. This equation agrees with (the gradient of) (1.7) in the special case that $\vec{u} = -\vec{\bigtriangledown}\Phi$ and $\dot{\vec{X}} = 0$. It also agrees with (1.6) in the complimentary case, when $\vec{\bigtriangledown}\Phi = 0$ and $\vec{u} = \dot{\vec{X}}$.

However, the combination that is normally used for Couette flow - Eq.(1.4) or Eq.(2.5) below - is a linear combination of solutions from two different theories!

Stationary flow is possible only if the effect of viscosity is negligible, this requires that $\bar{\mu} = 0$ or else that

$$\Delta\vec{u} = 0. \tag{2.4}$$

When the flow is stationary Eq.s (2.3)-(2.4) allow the horizontal flow

$$\vec{u} = \frac{a}{r^2}(-y, x, 0) + b(-y, x, 0), \quad a, b \text{ constant}, \tag{2.5}$$

with parameters $a, b$ to be determined by the non-slip boundary conditions. Both of these types of flow are of great interest, especially so in connection with superfluid Helium. [5]

This solution of the Navier-Stokes equation is a great success, so far, being in very good agreement with experiments, at low speeds of rotation.

**The difficulty**

is that, when both $a$ and $b$ are non-zero the two terms in (2.5) are solutions of two very different theories; so far, there has been no dynamical theory that is solved by the vector field (2.5) ($ab \neq 0$). The *ansatz* (2.5) is incompatible with either version of conservative hydrodynamics.

**One can not, in general, associate the Navier-Stokes equation with a "total velocity" $\vec{u}$, with an energy density $E[\rho, \vec{u}]$ or a kinetic potential $K[\vec{u}]$.**

This conclusion is long overdue and a turning point. The Navier-Stokes equation by itself does not imply, and in general it does not allow, the existence of an expression with the attributes of energy.

**Rayleigh's work and what it tells us**

The main challenge to theory has been to understand the limits on the stability of the laminar flow that was described above and the first notable attempt to do so was that of Rayleigh (1889, 1916) who concluded that the laminar flow should be stable if and only if

$$\omega_o r_o > \omega_i r_i > 0 \quad \text{(approximately)}.$$

Here $\omega_i(\omega_o)$ is the angular velocity of the outer (inner) cylinder and $\omega_i$ is positive by convention. This result is correct in the case that both cylinders rotate in the same sense, when the two angular velocities have the same sign. But it wrongly implies that the laminar flow would be unstable whenever the two cylinders rotate in opposite directions. The result of observation was different: the condition for stable Couette flow is, approximately, $\omega_i r_i > |\omega_o| r_o$, which allows for stable flow for either sign of $\omega_o/\omega_i$. Rayleigh's reaches his conclusion twice.

[5] This is exactly what is done in modern treatments; Rayleigh assumed that the viscosity is zero; consequently he cannot invoke the condition (2.4).

1. *Angular momentum (about the axis of rotation) is conserved ...* .

2. *We may found our argument upon a consideration of the kinetic energy...*
. In both cases, the expressions chosen by Rayleigh are *ad hoc* since he does not have an action, or any dynamical theory within which the 'energy' or the 'kinetic potential' can be defined, or shown to merit the title.

That Rayleigh obtained an unsatisfactory result is not significant because he did not have a well defined theory; lacking directions he was reduced to making *ad hoc* assumptions; that is our main conclusion.

The Navier-Stokes equation does not imply, and in general it does not allow, the existence of an expression with the attributes of energy. But a kinetic potential that is compatible with the Navier-Stokes equation exists in the special case of stationary, laminar Couette flow, when the velocity field is of the form (2.5); it is [6]

$$-K = \frac{b^2}{2}r^2 + ab\ln r^2 - \frac{a^2}{2r^2}, \qquad (\vec{u}\cdot\vec{\bigtriangledown})\vec{u} = \vec{\bigtriangledown}K. \tag{2.6}$$

By the Navier-Stokes equation, this makes a contribution $-\vec{\bigtriangledown}K$ to the acceleration. This is the only kinetic potential that is consistent with the Navier-Stokes equation in the simplest case when the velocity is of the form (2.5).

During the 100 years that followed the publication of Rayleigh's paper the calculation has been repeated in numerous textbooks. [7] That Rayleigh's prediction was contradicted by experiments must have been known to himself in 1916; that it was known to the other authors mentioned is not in doubt. Yet there is no suggestion in the later literature (that includes text books!) that Rayleigh's argument is invalid. An opportunity to draw more useful conclusions from the experiments was missed.

[6]The discovery of the signs in this expression set the direction of our research for more than 5 years (Fronsdal 2016).

[7]Including these: Chandrasekhar (1955 and 1980), Landau and Lifshitz (1959), Drazin and Ried (1981), Tilley and Tilley (1986), Koshmieder (1993) and Wikipedia (2017).

## III. Conservative hydrodynamics and the kinetic potential

The Hamiltonian density associated with the proposed Lagrangian (1.5) is

$$h = \rho\left(\dot{\vec{X}}^2/2 + (\vec{\bigtriangledown}\Phi)^2/2\right) + f. \tag{3.1}$$

It is, simply, the sum of the Hamiltonians of the two standard variational principles, the 'Eulerian' and the 'Lagrangian' versions. The equations of motion include the equation of continuity and the Bernoulli equation (Bernoulli 1738), obtained by taking the gradient of (1.7),

$$\vec{\bigtriangledown}(\dot\Phi - K - \varphi) = \frac{1}{\rho}\vec{\bigtriangledown}p, \tag{3.2}$$

$$K := -\dot{\vec{X}}^2/2 - \kappa\rho\dot{\vec{X}}\cdot\vec{\bigtriangledown}\Phi + (\vec{\bigtriangledown}\Phi)^2/2. \tag{3.3}$$

where $p$ is the thermodynamic pressure field in the bulk of the fluid.

The equations of motion include the Euler Lagrange equations and a gauge constraint, derived as usual by variation of the complete gauge theory Lagrangian with respect to the gauge field $Y_{0i}$ (Appendix) ,

$$\vec{\bigtriangledown}\wedge\vec{m} = 0, \qquad \vec{m} := \rho\dot{\vec{X}} + \kappa\vec{\bigtriangledown}\phi. \tag{A.1}$$

It is solved by

$$\vec{m} = -\vec{\bigtriangledown}\,\tau, \tag{3.4}$$

where $\tau$ is an arbitrary scalar density. Please see Appendix for details. Variation of the action with respect to the vector potential $\vec{X}$ gives $\frac{d}{dt}\vec{m} = 0$.

### Viscocity and stationary flows

In the presence of viscosity there can be no energy conservation and no action principle. But there is a standard and perhaps natural way to modify the Euler-Lagrange equations. Instead of (3.5), $d\vec{m}/dt = \bar\mu\rho\Delta\vec{v}$. The implication is that, for the flow to be stationary, the viscosity must be negligible; either $\bar\mu$ $= 0$ or $\vec{v}$ must be harmonic.

We have limited our attention to stationary flows with $\Delta\vec{v} = 0$, taking our inspiration from the traditional point of view. [8]

[8] We do not acually make use of Eq.(3.6). Of all possible flows those that are less affected by viscosity are favored to last. We assume, in agreement with standard theory, that they include harmonic flows. But certain types of energy loss do make contribution to the right hand side of Eq.(3.4).

## Stability

The simplest flows are horizontal and along circles centered on the cylinder axis, with $\Phi$ and $\tau$ proportional to the angular variable $\theta$,

$$-\vec{\bigtriangledown}\Phi = \frac{a}{r^2}(-y,x,0), \quad \vec{m} = -\vec{\bigtriangledown}\tau = \frac{b\kappa}{r^2}(-y,x,0). \tag{3.5}$$

The second formula solves the gauge constraint (A.1). The velocity of mass transport is

$$\vec{v} = \kappa\frac{\vec{m}}{\rho} - (1+\kappa^2)\vec{\bigtriangledown}\Phi = \left(\frac{\kappa^2 b}{r^2\rho} + (1+\kappa^2)\frac{a}{r^2}\right)(-y,x,0) \tag{3.6}$$

The only additional constraint on the field $\vec{v}$ is that it must be harmonic; that is, the density profile must take the form

$$\frac{\rho_o}{\rho} = 1 + \alpha\left(1 - \frac{r^2}{r_o^2}\right), \qquad \alpha \text{ constant}, \tag{3.7}$$

We shall learn, from the experiment, that $\alpha$ is related to $\kappa$, and that it is negative, contrary to what was expected. In terms of the parameters $a, b$,

$$-K = \frac{\vec{m}^2}{2\rho^2} - (\kappa^2+1)\vec{\bigtriangledown}\Phi^2/2 = \frac{1}{2r^2}\left(\frac{b^2\kappa^2}{\rho^2} - (1+\kappa^2)a^2\right). \tag{3.8}$$

## Boundary conditions

The walls of the two cylinders move with angular velocities

$$\omega_i\hat{\theta} = \frac{\omega_i}{r}(-y,x,0), \quad \omega_o\hat{\theta} = \frac{\omega_o}{r}(-y,x,0) \quad \omega_i, \omega_o \text{ constant}. \tag{3.9}$$

The velocity of mass transport is $\vec{v}$, Eq.(3.6), this is the velocity that must satisfy the no-slip boundary conditions, whence

$$\frac{1+\kappa^2}{r_i^2}(a + \frac{bc^2}{\rho_i}) = \omega_i, \quad \frac{1+\kappa^2}{r_0^2}(a + \frac{bc^2}{\rho_o}) = \omega_o, \tag{3.10}$$

and, if $\rho_o \neq \rho_i$,

$$a = \frac{r_o^2\rho_o\omega_o - r_i^2\rho_i\omega_i}{(\rho_o-\rho_i)(\kappa^2+1)}, \qquad b = \rho_o\rho_i\frac{r_i^2\omega_i - r_o^2\omega_o}{\kappa^2(\rho_o-\rho_i)}. \tag{3.11}$$

With this the preparations are complete and the experiments can proceed.

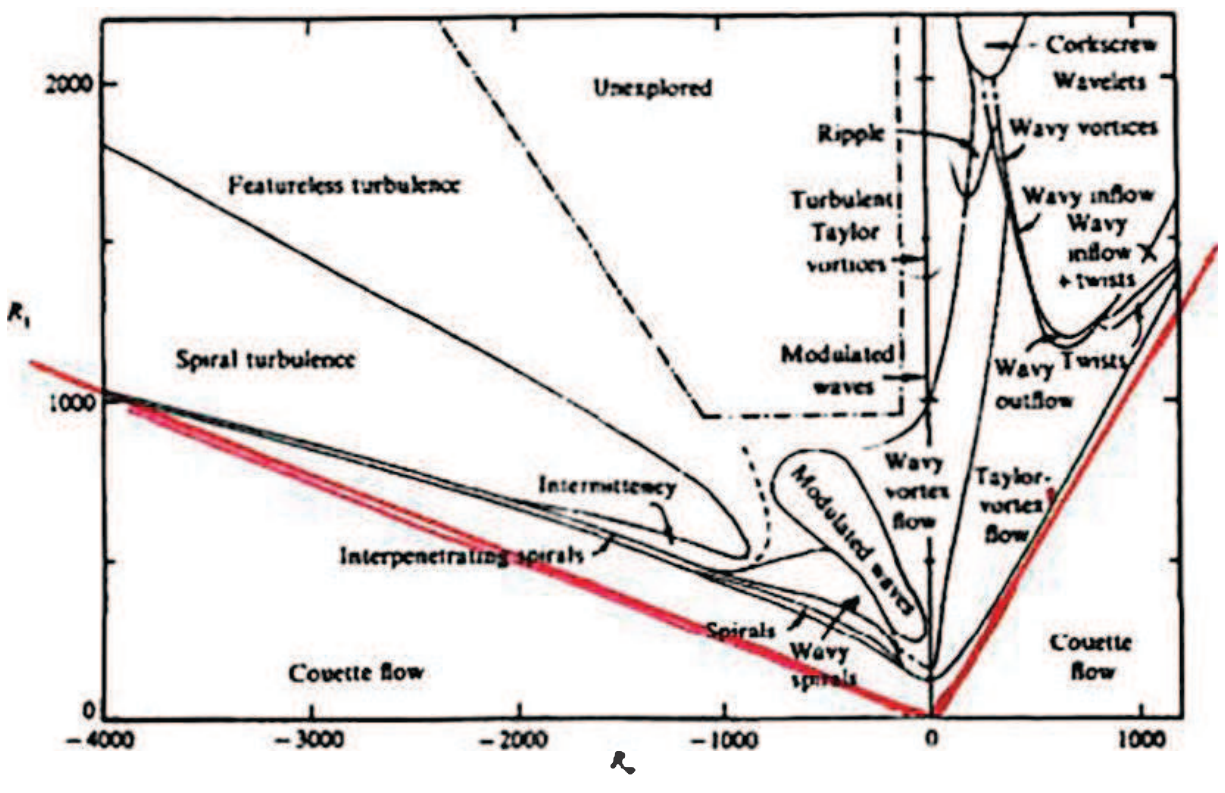

Fig.4. Experimental results of Andereck et al (1983, 1986). The abscissa (ordinate) is the angular velocity of the outer (inner) boundary. The lower 'hyperbola' is the upper limit of stability of laminar flow.

## IV. Experimental results and calculations

The experiments by Couette, Mallock and Taylor have been quoted already. We shall focus on results reported by Andereck et al, summarized in Fig. 4.

We have a legitimate kinetic potential and we can try Rayleigh's stability criterion; that is, that the lowest flow regime is stable only when $K' := \partial_r K(r) < 0$. But in this theory, with the density profile nearly uniform, the function $K'$ is negative almost everywhere; see Fig 5.

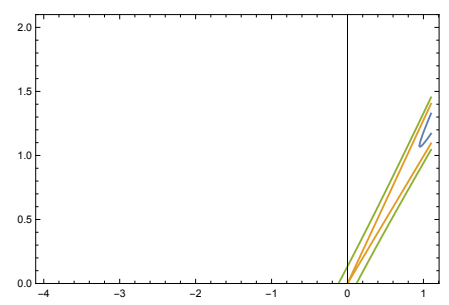

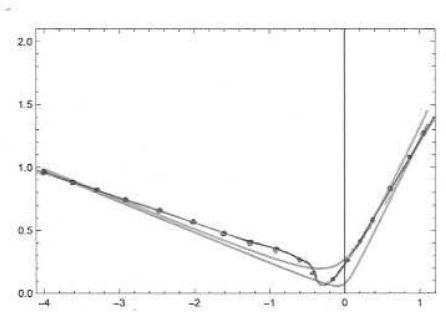

Fig.5. Typical $K'$ loci in the $(\omega_0, \omega_i)$ plane for $r = r_i$. This function is negative everywhere except between the two red lines through the origin.

Fig.6. Loci of $K(r_i)$, to illustrate the quality of the fit. The dotted curve is taken from Fig.4. For any assumed value of $\rho_i$ we choose the best value of $\kappa$.

Loci of $K(r)$ on the other hand strongly resemble the limit line in Fig.4 – as illustrated in in Fig.s 6 and 7 – provided that the parameters $\alpha$ and $\kappa$ are chosen so that the boundary values satisfy $\rho(r_i) > \rho(r_o)$. This result is initially somewhat anti-intuitive and it is an interesting prediction of the theory. There have been some measurements of the local density in the Couette system using (gaseous) Argon, but we found not a single one using liquids.

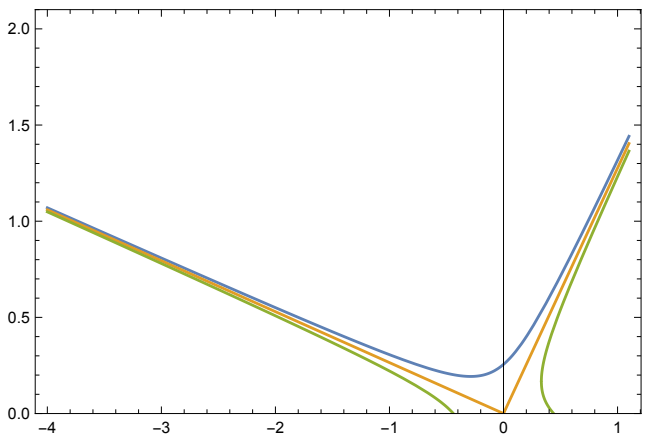

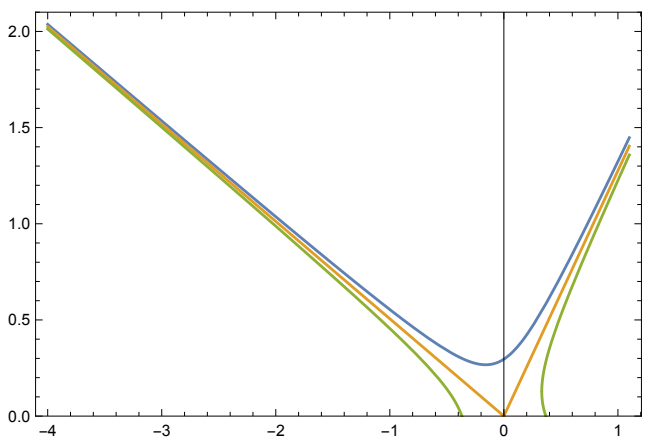

Fig.7. $K(r_i)$ loci. The upper (blue) curve is a best fit to the stability limit at the inner boundary. As the value of $r$ moves outwards, the left branch of the limit moves uniformly clock-wards in the $\vec{\omega}$- plane, while the right branch rotates very slightly in the other direction.

Fig.8. $K(r)$ loci at $r/r_o = .92$.

## Proposed stability criterion

It is known that some types of flow instability are accompanied by bubble formation, as in the wake of propellers, and in the case of the rupture in experiments that produce metastable states. A popular interpretation of sonoluminescence postulates that it is mediated by the creation and subsequent collapse of tiny bubbles. This suggests that the limit of stability is marked by a 'critical' value of $K$. Since the value of $K$ is strongly correlated with the value of the pressure, this means that the instability occurs at a point that is determined by standard thermodynamics, by the value of the pressure or the value of the density. Experiments on metastable configurations of water with negative temperatures, by Azouzi *et al* (2012) and others confirm that rupture proceeds by cavication, wich suggests that the critical pont is the thermodynamical bubble point.

The experimental evidence is complicated, especially in view of important end effects. We quote Andereck *et al* (1986): "If $R_o < -155$ ... the basic flow is centrifugally unstable to non-axisymmetric spiral vortices near the inner cylinder." The region SPI is near $\vec{\omega}$ = (-300,300).

For a first trial we set $\rho_o/\rho_i = .9$ and adjusted the parameter $\kappa$. Fig.6 illustrates the quality of the fits. Fig 7 shows a $K(r_i)$ locus (top line) that best fits the experimental curve with $\rho_o/\rho_i = .9,\ \kappa = 5.3$. Very similar loci were found for the parameters satify the empirical relationship

$$\rho_i/\rho_o - 1 \approx \frac{3}{\kappa^2}, \quad 5 < \kappa < 500. \tag{4.1}$$

This condition has to be satisfied in order for the theory to agree with the experiments in the case of counter-rotating cylinders. In the co-rotating case it can be relaxed; this case will be discussed in Section 5.

The slope of the right asymptote is predicted without ambiguity to be close to $r_o/r_i$. The slope on the left is very sensitive to the value of $\alpha$.

For negative values of $\omega_o$, as we move outwards, to larger values of $r$, the limiting locus of $K(r)$ rotates uniformly clockwise, confirming that the breakdown of stability begins (as $\omega_i$ is increased) at the inner boundary. As this suggests that breakdown can be observed only at the inner boundary we show only one other example, at $r = .92$, in Fig.8.

The next two illustrations show flow lines. They give the impression that the liquid is being dragged along by the volume elements that are holding on to the walls (no-slip boundary condition), and perhaps this suggests an increased pressure near the inner cylinder.

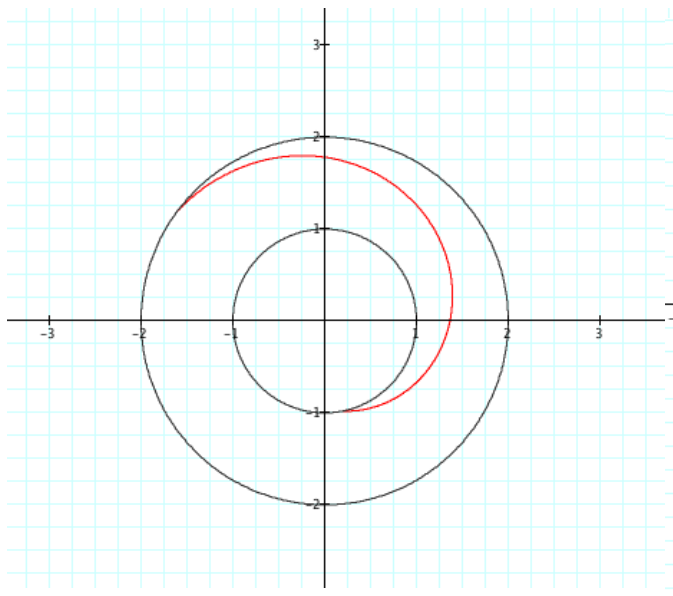
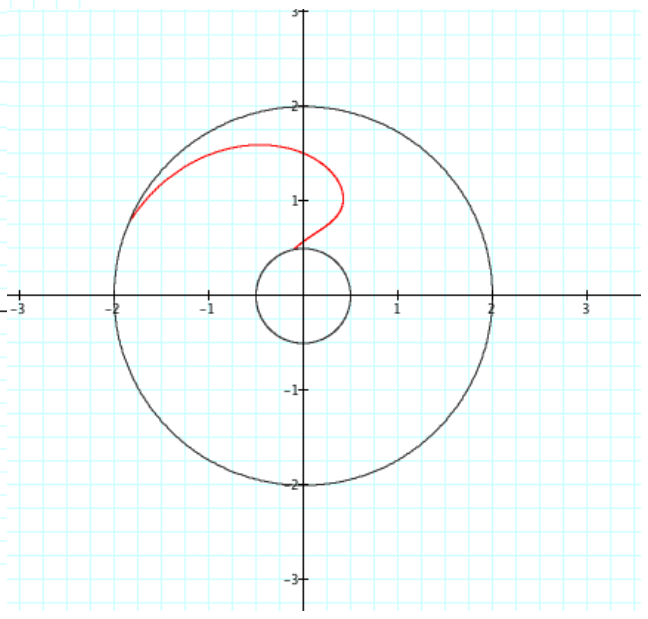

Fig.9. Example of flow lines, co-rotation, counter clockwise flow.

Fig.10. Example of a flow line, contra-rotation, counter clockwise flow.

The adiabatic compressibility of water under normal conditions has the (negative) value

$$\beta = \frac{\partial \ln V}{\partial p} = -.25\ 10^{-4},$$

where $V$ is the adiabatic, specific volume.

In our model,

$$V = \rho_o/\rho = 1 - \alpha(x^2 - 1). \quad x := r/r_o. \tag{4.2}$$

Since $\alpha$ is negative, $V$ and $p$ both increase outwards, to make $\beta$ positive, which is highly unexpected. With $\kappa = 5.3$ it takes the value .25 $10^{-4}$, which makes it unstable by classical stability theory. In this geometry the fluid is being compressed in the radial direction and stretched in the angular direction; we do not have an intuitive understanding that could help us evaluate this result.

The discovery that reality is different from standard theory is not unfamiliar in hydrodynamics; recall the radiometer and similar phenomena.

**Numerical predictions, counter rotation.**

Unit conversion. The basic unit of angular velocity is 1 radian per second. The units used in Fig.4 are $R_o = r_o(r_o - r_i)\nu = 414\ rad/sec$ and $R_i = r_o(r_o - r_i)\nu = 365.4\ rad/sec$, where $\nu$ is the viscosity. We direct our attention at the critical point [9]

$$(R_o, R_i) = 3000, 900), \quad (\omega_o, \omega_i = (-7.25, 2.47). \tag{4.3}$$

By way of example we set $\kappa = 5.326$. We have explored widely and the conclusion is that the value of $\kappa$ can be changed from 1 to 1000 without any radical change in the landscape. The other parameters are much less flexible. The critical value of $K$ is near zero, the numerical coefficient in Eq. (4.1) can be changed by 10%.

With these values we find that

$$a = 102.7. \quad b = -115.4. \tag{4.4}$$

We used these values of $a$ and $b$ to calculate $v$ as a function of $r$, plotted in Fig.11. This is a theoretical interpolation between endpoint values realized in the experiment, assuming that the no-slip boundary conditions were valid.

In Table 1 we have listed, for the chosen values of $R_o$ and $R_i$, in the first row (labelled $v$), the range of values of this variable for $r_i < r < r_o$. In the second row we have listed values of the function $-\tilde{K}' := v^2/r$. This function is of interest because it is commonly assumed that it is an effective kinetic potential; hence the notation. The function $\rho v^2/2$, the traditional formula for energy density (not shown) has a very similar shape over the short range under consideration. In remarkable contrast, the Hamiltonian energy density $h$ (Table 1 and Fig.11) shows an almost flat distribution. Next we calculate the kinetic potential $K(r)$, defined in Eq.(3.9), in cgs units, $cm^2/sec^2$. We note that the kinetic potential is negative and decreasing outwards.

[9] The stable point $(R_o, R_i) = (-3000, 500)$ is at$(\omega_o, \omega_i) = (-7.25, 1.37)$.

At the critical point (4.3) the acceleration is

$$v_o^2/r_o = r_o\omega_o^2 = 400cm/sec^2 = .4g.$$

The pressure $p_o$ - at this point is "probably" 1.4 atm, the value obtained by adding the last number, to the ambient value of the gradient of the pressure. To confirm this we use the Bernoulli equation, in the form

$$K + gz + \mu = \text{constant}.$$

The chemical potential $\mu$ is related to the pressure,

$$\mu' = p'/\rho.$$

In this calculation it is acceptable to treat the density as a constant, then

$$K + gz + p/\rho = \text{constant} \tag{4.5}$$

and

$$p_i/\rho_i + K_i = p_o/\rho_o + K_o.$$

At the walls

$$K_o = -950, \quad K_o' = -1261 = -1.2g, \quad K_i = 55, \quad K_i' = -1360 = -1.36g.$$

This makes the pressure at the inner wall very close to 0, which is near the pressure of the bubble point of water at normal temperature.

We calculated $p(r)$ at the chosen point and plotted the pressure as a function of the density, shown in Fig 12.

By the way, the adiabatic derivative of the pressure is negative, as is normal in this domain that, in the absence of motion, is metastable. What is highly unexpected is that the density decreases outwards. It can be understood as a natural effect of tension.

In the case of opposite rotation the mass flow velocity changed sign near the inner boundary. Here the physics seems to be very similar to the phenomenon of negative pressure in Berthelot tubes, where negative pressures can be realized, and capillary action. [10], which brings back to mind the interesting parallel between the present theory of two flows and Landau's preferred model of phonons and rotons) over the competing model of 2 fluids.)

Table 1: Couette profiles. Results for $\kappa = 5.3$. Numerical profiles for $\omega_o - 3000, \omega_i = 0, 500$ and 900, $\rho_o/\rho_i = .9$.

| | $1/\rho_i = .9$ | $1/\rho_i = .9$ | $1/\rho_i = .9$ |
|---|---|---|---|
| $\vec{\omega}$ | (-7.25, 0) | (-7.25, 1.37) | (-7.25, 2.47) |
| $v$ | 0 to -43.1 | 8.0 to -43.1 | 14.2 to -43.1 |
| $v^2/r$ | 0 to 312 | 10.7 to 312.5 | 34.9 to 312 |
| $h$ | 112 to 141 | 148 to 175 | 192 to 214 |
| $K$ | -112 to -900 | -34.3 to - 1061 | 64.7 to -1196 |

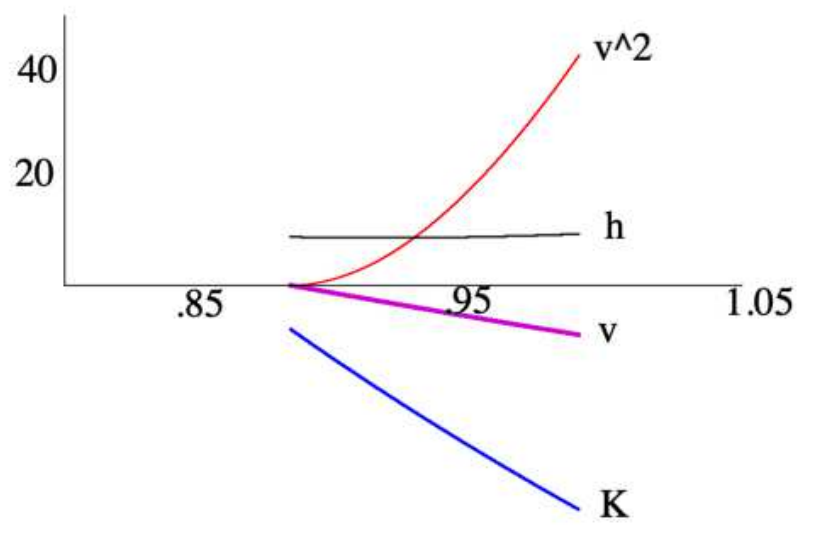

Fig.11. Profiles of speed, energy, kinetic energy and pressure ($p = -\rho K$).

[10]One implication is that the tension can explain the remarkable capture of angular momentum seen in certain experiments with frozen superfluid Helium (Kim and Chan 1950

**Summary, contra-rotation: The theory predicts that, when the instability sets in, the pressure falls to near zero at the inner boundary. Local measurements of the pressure would be of the greatest interest.**

# V. Co-rotating cylinders

The physics that is revealed by a study of the contra-rotating case is of interest because it relates to the phenomena of negative pressures and the existence of liquid tension. The case of co-rotation, when the cylinders rotate in the same direction is very different. There are 2 parameters to vary. the parameter $\kappa$ is characteristic of the fluid and is therefore the same for both cases, con- and contra-rotation. In co-rotation we find no tendency towards an important reduction of the pressure; the phenomenon that dominates the other case does not happen here. With a less important variation of the pressure we expect a smaller variation of density. The parameter $\rho_i/\rho_o$ is therefore expected to be quite close to unity.

With $\kappa$ in the range 3-10 and $\lambda := (\frac{\rho_i}{\rho_o} - 1)\kappa^2$. of the order $10^{-5} - 10^{-6}$ we fond that $K'$ turns positive as $\omega_i/\omega_o$ increases beyond the value $(r_o/r_i)^2$, as predicted by Rayleigh and in agreement with the experiment by Andereck et all. Apparently, there is no need to invoke the two-vector theory in the case of co-rotation. This observation is analogous to what is seen in the experiments with negative pressures.

**No central cylinder.**

The remarkable degree to which the observed line of instability agrees with a locus of $K$ supports the notion that other predictions of the theory may be relevant. Of particular interest are the flow and density profiles in the case that the inner cylinder is removed. In the limiting case when the radius of the inner cylinder tends to zero the no - slip boundary condition makes sense if

$$\frac{b}{a} = -(1 + \kappa^{-2})\rho_i/\rho_o \tag{5.1}$$

and in that case $\vec{v} = 0$ at the center - see Eq.(3.8). In other words, the "solid body" part of the flow regularizes the irrotational part of mass flow at the center, just as was explained by Onsager back in the 1950's. This desireable result is due to the fact that the fluid is compressible, more precisely to the factor $\rho$ in Eq.(1.9 bis).

Even if $\vec{v} = 0$, $K(r)$ is still singular at the center; so long as $\Phi \neq 0$. This must be so, for the the kinetic potential $\Phi$ in $K \propto \vec{\bigtriangledown}\Phi^2$ is what holds up the meniscus on a liquid at rest in the gravitational field.

All this is possible because the density is not uniform,

$$\frac{\rho_o}{\rho} = \Big(1 + \alpha(1 - r^2/r_o^2)\Big)^{-1}, \quad \alpha = \text{constant},$$

which was derived from the harmonicity of $\vec{v}$. With a value of $\alpha < 1$ the density is very nearly constant except near the center and the velocity of flow is very nearly constant, as found in older experiments at low rotation speeds (Lamb 1932, p. 1). The fact that it is not quite constant is interesting and should be checked.

# VI. Conclusions

Energy, with associated conservation laws, is the very soul of theoretical physics. It is natural that, in the many contexts where an expression for it is not available, attempts are made to invoke it anyway. But a more ambitious goal is to formulate all of physics in terms of action principles. This does much more than provide the needed expressions for the components of the energy momentum tensor, for it improves the power of prediction to a high degree.

To understand real fluids it is necessary to recognize that all fluids are compressible. This is dramatically confirmed by the first prediction of the present theory: the vorticity is perpendicular to the gradient of the density; Eq.s, (1.7) and (1.8) [11] This should be relatively easy to verify (or refute) experimentally.

Another prediction that stands out is that, in contra-rotating Couette flow, the pressure falls to near zero (at the inner wall) when angular velocity is increased to the limit of stability of the most basic flow. This has not been verified in the classical version of Couette flow, but a related experiment (Ref.) is determinant. A tube bent into the form of a Z is filled with water and rotated about a central, vertical axis. As expected, the centrifugal force causes a reduction of the pressure at the axis, in the middle of the inclined part of the tube. The flow is stable until a vacuum bubble forms at this point. The pressure is thus at the bubble point at the given temperature. Other experimenters have reached the same conclusion, for example Azouzi *et al* (2012).

A viable action principle for rotational Hydrodynamics, Thermodynamics and General Relativity has been proposed. It is not unique, for complicated systems require more variables, but it is the most economical. This paper offers a first instance of an application, to a system that has resisted a complete development even in the limit of incompressible flows.

Additional support for the present interpretation of the new vorticity field comes from a study of the internal dynamics of menisci and the metastable states of negative pressure in Berthelot's experiment.. (Fronsdal 2020c) in preparation.)

[11] In a larger context the prediction is made when magnetic fields are absent or irrelevant.

The lifting of this model to General Relativity is well known (summarized in Fronsdal 2019). An application to planetary systems, in the Newtonian approximation, is in preparation. [12]

**A wider perspective**

Some or all instabilities in (traditional, equlibrium) thermodynamics are associated with phase transformations, but so far there is no theory of phase changes in the borderland of adiabatic (non equilibrium) thermodynamics and hydrodynamics. The phenomenon studied in the paper is a stationary process that belongs as much to Adiabatic Thermodynamics, a subject first explored by Newton and Laplace in their calculation of the speed of sound and 2-velocity hydrodynamics. [13]

Adiabatic thermodynamics here refers to a theory defined by an action principle. The action, when applied to configurations of equilibrium, defines traditional thermodynamics. In the case of simple systems in motion the equations of motion are those established by Callen for "extended thermodynamics". When the temperature is eliminated by means of the adiabatic condition one obtains the equation of continuity and the Bernoulli equation with a potential that depends on the entropy.

For a fluid with specific free energy function $F$ the entropy is defined by the Euler-Lagrange equation:

$$\frac{\partial F}{\partial V} + S = 0,$$

also known as the adiabatic relation. After the value of $S$ has been fixed this equation is a constraint that allows to eliminate the temperature in favor of $S$. The specific entropy $S$ is always, in this context, taken to be constant, an adiabatic invariant.

Incidentally, this equation predicts a temperature profile for Couette flow. We are not aware of any measurements aimed at verifying it.

There is a well known and eminently successful theory of the phase transformation in a static van der Waals fluid, in particular, evaporation/condensation of water. We propose extending this theory to water in a (nearly) stationary state of rotation. Water in irrotational motion has 2 independent variables and the free energy density is replaced by

$$h(\rho, \Phi) = \rho\vec{\bigtriangledown}\Phi^2/2 + f(\rho),$$

where $f$ is the free energy density of a static van der Waals' fluid.

$$f = \rho\mathcal{R}T\ln(\frac{\rho}{(1-b\rho)T^n} - a\rho^2.$$

[12]There is a preliminary version in arXiv

[13]for a treatment that sees the problem almost entirely as one of thermodynamics see Debenedetti, 1996.

The chemical potentials are

$$\partial\tilde{f}/\rho = RT(\ln\frac{\rho}{(1-b\rho)T^n} + \frac{1}{1-b\rho}) - 2a\rho + v^2/2, \tag{6.1}$$

$$\partial h/\partial\Phi = -\vec{\bigtriangledown}\cdot(\rho\vec{\bigtriangledown}\Phi). \tag{6.2}$$

Continuity of this second chemical potential is assured by the equation of continuity.

The next step is to calculate the points of coexistence of two phases, the dew point (a) and the bubble point (b) where $T, p$ and $\vec{\mu}$ have the same values, $T(a) = T(b), p(a) = p(b), \vec{\mu}(a) = \vec{\mu}(b)$. Here $\vec{\mu}$ is the 2-component chemical potential,

$$\vec{\mu}(\rho,\Phi) = \vec{\bigtriangledown}\tilde{f} := (\partial\tilde{f}/\partial\rho, \partial\tilde{f}/\partial\Phi).$$

In addition, we need to expand our horizons in other directions. Andereck and others have explored the full range of flow regimes in the Couette problem; we need to vary the fluid (even as far as using gases) as well as the density of absorbed air, ambient temperature and pressure and, especially, surface tension; that is, the preparation of the surfaces of the cylinders.

## Predictive power.

The acid test of a theory in physics is predictive power. With a history of physics spanning 400 years and more it is easy to see what it is that gives a theory this property:

**Symmetry.** Apart from esthetic appeal, symmetry is favored everywhere because it unifies different phenomena in a framework that implies a large reduction of arbitrariness: energy is preserved, hence time independent, masses are equal. Under this rubrick comes relativistic invariance.

**Action principles**, an example of structure and economy of expression. With Poicaré we feel that a set of equations thrown togeter needs a soul to be convincing; in most successful theories the soul is an Action that in a single expression is the source of all the equations of motion.

Internal consistency such as unitarity and renormalizability.

In a search for high predictive power the preferred model of hydrodynamics was Lagrange's action principle for irrotational flow. This theory gave evidence of high preditive power, including the prediction that flight is impossible. Important phenomena have been explained after discovery but none have been predicted. Usually, the action was discarded, in the case of Lagrange's theory this leaves a theory that consists of two equations, the equation of continuity and the Bernoulli for $\rho$ and a general vector field $\vec{v}$. But to use the Bernoulli equation we need the expression for the kinetic potential.

**Unitarity.** There have been many suggestions to the effect that unitarity, a crucial concept in quantum theories, has an analogue in classical field theories. But, as far as we know this lead has not yet been followed in the context of hydrodynamics.

a. We need a generalization of irrotational flow. This cannot be done with only scalar fields.

b. Coordinate transformations play an important role. It is a group that acts on fields including the vector field $\vec{X}$, in standard fashion. The equation

$$\vec{\bigtriangledown} \cdot \vec{X} = 0. \tag{6.3}$$

is invariant under this action. We can restrict ourselves to the solutions of this equation, or to the complement of this subspace. Another ivariant equation is

$$\vec{\bigtriangledown} \wedge \vec{X} = 0. \tag{6.4}$$

Again. we can restrict ourselves to the subspace of vector fields that satisfy this equation, or to the complement.

This gives us 5 alternatives: select one out of 4 subspaces, or impose neither of the conditions. But what is the restriction imposed by "unitarity"?

It is not known to the author whether this programme has been carried out. But in the expectation that any relevant physical theory must have a relativistic generalization the answer is known: (6.3) leads to electromagnetism and (6.4) to the 2-form gauge theory of Ogievetskij and Polubarinov with only 2 degrees of freedom. This gives us a second theory of hydrodynamics and, in particular, a dynamical realization of Landau's two-flow approach to superfluids. It has often been said that superfluid Helium is unique only because all other fluids solidify at low temperatures. Consequetly it is natural to apply the theory to other fluids, including water, as we have done.

**References**

Andereck, C.D., Dickman, R. and Swinney, H.L."New flows in a circular Couette system with co-rotating cylinders", Physics of Fluids, **26**, 1395 (1983); doi 10.1063/1.864328.

Andereck, C.D., Liu, S.S. and Swinney, H.L."Flow regimes in a circular Couette system with independently rotating cylinders", J.Fluid Mech. **164** 155-183 (1986).

Azouzi, M., Ramboz, C., Lenain, J. et al. A coherent picture of water at extreme negative pressure. Nature Phys 9, 3841 (2013).

Bernoulli, D., *Argentorat*, 1738.

Berthelot, M., Ann. chim. Physs. 30, 232-237 (1850)

Betchov, R. and Criminale, W. O. J., Stability of parallel flows, New York,Academic Press, 1967.

Birkhof, G. *Hydrodynamics*, Princeton U. Press (1950)

Chandrasekhar, S., *Hydrodynamics and hydrodynamic stability*, Oxford Univ. Press (1961).

Couette, M., "Oscillations tournantes d'un solide de révolution en contact avec un fluide visqueux," Compt. Rend. Acad. Sci. Pari**105**, 1064-1067 (1887).

Drazin, P.G. and Reid, W.H., , *Hydrodynamic Stability*, Cambridge University Press (1981)

Eckert, (2006). "The Beginnings of Fluid Mechanics in Goottingen, 1904-19014. The Dawn of Fluid Dynamics: A Discipline Between Science and Technology. Weinheim.

Einstein, A. 1915 " Ideas of the General Theory of Relativity and the Application of this Theory in Astronomy", Preussische Akademie der Wissenschaften, Sitzungsberichte, 1915 (part 1), 315

Fronsdal, C. "Ideal stars in General Relativity", Gen. Rel. Grav **39**, 1971-2000 (2007)

Fronsdal, C. , "Heat and Gravitation. The action Principle", Entropy 16(3), 1515-1546 (2011)

Fronsdal, C., "Relativistic Thermodynamics, a Lagrangian Field Theory for general flows including rotation", Int.J. Geom.Math.Meth. Phys., (2016)

Fronsdal, C., "Action Principles for Thermodynamics, Review and Prospects", Phys. Elem. Part. and Nucl. (JINR) (2017a).

Fronsdal, C., "Action Principles for Thermodynamics" in *Memorial Volume on Abdus Salam's 90th Birthday*, Lars Brink, Michael Duff and Kok Khoo Phua (eds.), (World Scientific Publishing), ISBN: 978-981-3144-86-6 (2017b).

Fronsdal,C., "Sources for Gravity" arXiv (2020b)

Fronsdal, C. *Adiabatic Thermodynamics of Fluids. From Hydrodynamics to General Relativity*, monograph, World Scientific (2020a)

Fronsdal' C., "Below the Meniscus", (2020b)

Fronsdal,C., "Metastable fluids, capillary action and superfluids", in preparation (2020c )

Fronsdal. C., "Wind tunnels. The drag of a sphere." In preparation (2020d)

Hall, H.E. and Vinen, W.F., "The Rotation of Liquid Helium II. The Theory of Mutual Friction in Uniformly Rotating Helium II", Proc. R. Soc. Lond. A 1956 238, doi: 10.1098/rspa.1956.0215 (1956)

Kim, E. and Chan, M. “Probable observation of a supersolid helium phase’, Nature. 2004;427(6971):225227. doi:10.1038/nature02220
Koschmieder, E.L., *Benard cells and Taylor vortices*, Cambridge 1993.
Joseph, D. D., Stability of fluid motions, Vol.1 and 2, Berlin, Springer-Verlag, 1976.
Lagrange, J.M., Taurinensia, ii., (1760), Oeuvres, Paris, 1867-92.
Lagrange, J.M., ” Memoire sur la Théorie du Mouvement des Fluides ”, Notiv. mem. de VAcad. de Berlin, (1781)
Landau, L., “Theory of Superfluid Helium II”, Phys. Rev. **60** , 356-358 (1941)
Linus, F. “Tractatus de corporum inseparabilitate...”. (London 1661).
Lund,F and Regge, T., “Unified approach to strings and vortices with soliton solutions”, Phys.Rev. D **14**, 1524-
Lin, C. C., The theory of hydrodynamic stability, Cambridge, Cambridge Press, 1955.
Mallock, A., Proc.R.Soc. **45** 126 (1888).
Mallock, A., Ohilos.Trans.R.Soc. **187**, 41 (1896).
Munk, W.H. and Sverdrup, H.U., “Wind, sea ands well: theory of relations for wind forecasting”, Hydrographic Office, Boston (1947)
Ogievetskij, V.I. and I.V.Polubarinov, I.V.,“Minimal interactions between spin 0 and spin 1 fields”, J. Exptl. Theor. Phys. (USSR) **46**, 10481055 (1964)
Prandtl, see Eckert 1914.
Rayleigh, L., On the stability or instability of certain fluid motions, Proc. Lond. Maths. **11**, 1880, 57-70.
Rayleigh, L., “On the dynamics of revolving fluids”, Proc.Roy.Soc. A **93**148-154 (1916).
Taylor, G.I., “Stability of a viscous fluid contained between two rotating cylinders”, Phil.Trans. R.Soc. London **A102**, 644-667 (1923).
Tilley, D.R. and Tilley J., Superfluidity and Superconductivity, Adam Hilger Ltd, Bristol and Boston (1986)
von Neumann, J., Memorandum addressed to O. Veblen, in Collected Works, edited by A. H. Taub (Pergamon, New York,1963), page 98
b

# Appendix

Part 1. We present additional reasons to maintain, in a generalization of Eulerian hydrodynamics, the scalar velocity potential.

As important as the equation of continuity is the tradition that incorporates gravitation into field theories, especially into hydrodynamics, by including the Newtonian potential as a contribution to the Hamiltonian density. (Granted that it is suggested by the correspondence between hydrodynamics and particle mechanics, but our subject is field theories.) Newtonian hydrodynamics is a non-relativistic approximation to General Relativity, but is there actually a generally-relativistic field theory that has the expected non-relativistic limit? Yes, there is. Consider the relativistic field theory with action (Fronsdal 2007)

$$A = \frac{-1}{16\pi G}\int d^4x\sqrt{-g}R$$
$$+\int d^4x\sqrt{-g}\left(\frac{\rho}{2}(g^{\mu\nu}\psi_{,\mu}\psi_{,\nu}-c^2)-W[\rho]\right). \qquad (A.1)$$

Gravitation is represented by the metric, in the non relativistic limit by the time-time component. To explore the non-relativistic limit one sets

$$g_{00}=c^2+2\varphi,\quad g_{11}=g_{22}=g_{33}=-1,$$

off diagonal elements zero, expands (1.4) in powers of (1/c) and takes the limit $1/c\to 0$. To ensure the cancellation of the terms of lowest order we need to set

$$\psi=c^2t+\Phi,$$

then terms of order $c^2$ cancel and we are left with the action (1.1). In the context of field theories this is the best (perhaps the only) confirmation of the expectation that Non-Relativistic Hydrodynamics, with the inclusion of the gravitational potential, is the non-relativistic limit of General Relativity. Without the expansion $\psi=c^2t+\Phi$ the role of gravitation in hydrodynamics would loose the only direct contact that it has with General Relativity. It leads to the correct transformation law for the field $\Phi$ under Galilei transformations, given that $\psi$ is a scalar field on space time. But this theory is limited to irrotational flows.

Part 2. We summarize the constraints that characterize the relativistic gauge theory and the non relativistic Conservative Hydrodynamics.

The theory of a massless 2-form ($Y_{\mu\nu}$ was developed by Ogievetskij and I.V.Polubarinov (1964). It is invariant under the gauge transformations

$$\delta Y_{\mu\nu}=\partial_\mu\xi_\nu-\partial_\nu\xi_\mu$$

or $\delta Y=d\xi$. Tthe principal part of the relativistic Lagrangian is

$$\rho\frac{c^2}{2}\rho dY^2,$$

where the scalar density $\rho$ plays a role similar to the permittivity in electromagnetism. To this one can add

$$\epsilon dF^2 + \gamma \tilde{F}\tilde{Y},$$

where the second term massifies the photon, but for the present purpose it is important to include instead the gauge invariant interaction

$$\kappa \rho dY d\psi,$$

where $\psi$ is the scalar field that was introduced in (A.1).

The equations of motion include the Euler Lagrange equations and a gauge constraint, derived as usual by variation of the complete gauge theory Lagrangian with respect to the gauge field $Y_{0i}$,

$$\vec{\bigtriangledown} \wedge \vec{m} = 0, \qquad \vec{m} := \dot{\vec{X}} + \kappa \vec{\bigtriangledown}\phi.$$

It is solved by

$$\vec{m} = -\vec{\bigtriangledown}\tau.$$

Variation of the action with respect to the vector potential $\vec{X}$ gives

$$\frac{d}{dt}\vec{m} = 0.$$

Variation with respect to the gauge field gives the gauge constraint

$$\vec{\bigtriangledown} \wedge \vec{m} = 0.$$

Finally, the static sector is characterized by $\vec{\bigtriangledown} \cdot \vec{X} = 0$.